\documentclass[useAMS,usenatbib]{mn2e}

\usepackage{graphicx}
\usepackage{mathrsfs}
\usepackage{natbib}
\usepackage{amssymb}
\usepackage{overpic}
\bibpunct[,]{(}{)}{;}{a}{}{,}

\def\sun{{\odot}}

\newcommand{\begit}{\begin{itemize}}
\newcommand{\enit}{\end{itemize}}
\newcommand{\begen}{\begin{enumerate}}
\newcommand{\enen}{\end{enumerate}}

\newcommand       \be           {\begin{equation}}
\newcommand       \ee           {\end{equation}}
\newcommand       \bea          {\begin{eqnarray}}
\newcommand       \eea          {\end{eqnarray}}

\newcommand{\beqa}{\begin{eqnarray}}
\newcommand{\eeqa}{\end{eqnarray}}

\title[Diagnosing black hole formation with kilonovae]{Red or blue?  
A potential kilonova imprint of the delay until black hole formation following  a neutron star merger} 
\author[B.~D.~Metzger \& R. Fern\'andez]{Brian D. Metzger$^1$ and Rodrigo Fern\'andez$^{2,3}$\\
$^1$ Columbia Astrophysics Laboratory, Columbia University, New York, NY 10027, USA.\\
$^2$ Department of Physics, University of California, Berkeley, CA 94720, USA.\\
$^3$ Department of Astronomy \& Theoretical Astrophysics Center, University of California, Berkeley, CA 94720, USA.
}

\begin{document}

\date{Submitted to MNRAS}
\pagerange{\pageref{firstpage}--\pageref{lastpage}} 
\pubyear{2013}
\maketitle
\label{firstpage}

\begin{abstract}
Mergers of binary neutron stars (NSs) usually result in the formation of a
hypermassive neutron star (HMNS).  Whether- and when this remnant collapses to
a black hole (BH) depends primarily on the equation of state and on angular
momentum transport processes, both of which are uncertain.  Here we show that
the lifetime of the merger remnant may be directly imprinted in
the radioactively powered \emph{kilonova} emission following the merger.  We
employ axisymmetric, time-dependent hydrodynamic simulations of remnant
accretion disks orbiting a HMNS of variable lifetime, and characterize the
effect of this delay to BH formation on the disk wind ejecta.  
When BH formation is relatively prompt ($\lesssim 100$ ms), 
outflows from the disk are sufficiently neutron rich
to form heavy $r$-process elements, 
resulting in $\sim$week-long emission with a
spectral peak in the near-infrared (NIR), similar to that produced by the
dynamical ejecta.  In contrast, delayed BH formation 
allows neutrinos from the HMNS to raise the electron fraction in the polar direction
to values such that potentially \emph{Lanthanide-free}
outflows 
are generated.  The lower opacity would produce a brighter,
bluer, and shorter-lived $\sim$ day-long emission (a `blue bump') prior to the
late NIR peak from the dynamical ejecta and equatorial wind.  
This new diagnostic of BH formation should be
useful for events with a signal to noise lower than that required for direct
detection of gravitational waveform signatures.  
\end{abstract}


\maketitle

\section{Introduction}

Mergers of binary neutron stars (NSs) (hereafter `neutron star
mergers', or NSMs) are the primary source of gravitational waves (GW) for
upcoming ground-based interferometric detectors such as Advanced LIGO and Virgo
(\citealt{Abadie+10}).  They are also promising central engines for
short-duration gamma-ray bursts (GRBs; \citealt{Paczynski86,Eichler+89};
see \citealt{Berger13} for a recent review).

General relativistic simulations of NSMs 
show that the merger process can result in two
qualitatively different outcomes, depending primarily on the total mass of
the binary $M_{\rm t}$.  If $M_{\rm t}$ exceeds a critical value $M_{\rm c}$,
then the massive object produced by the merger collapses to a black hole (BH)
on the dynamical time ($\sim$few ~ms, e.g., \citealt{sekiguchi2011}).  
On the other hand, if $M_{\rm t} < M_{\rm c}$ then the merger product is at least
temporarily supported against gravitational collapse by differential rotation
and/or thermal pressure. 
This meta-stable compact object is usually called a \emph{hypermassive} 
NS (HMNS; e.g. \citealt{kaplan2013}).

The value of $M_{\rm c}$ depends on the uncertain equation of state of (EoS) of
nuclear matter. 
The recent discovery of massive
$\sim 2M_{\odot}$ NSs (\citealt{Demorest+10,Antoniadis+13}) 
excludes a soft EoS, 
placing a lower limit of $M_{\rm c} \gtrsim 2.6-2.8M_{\odot}$
(\citealt{Hotokezaka+13}; \citealt{Bauswein+13}).  It thus appears likely that the `canonical' 1.4 +
1.4 $M_{\odot}$ binary merger goes through a HMNS phase.

\begin{figure*}
\includegraphics*[width=1.5\columnwidth]{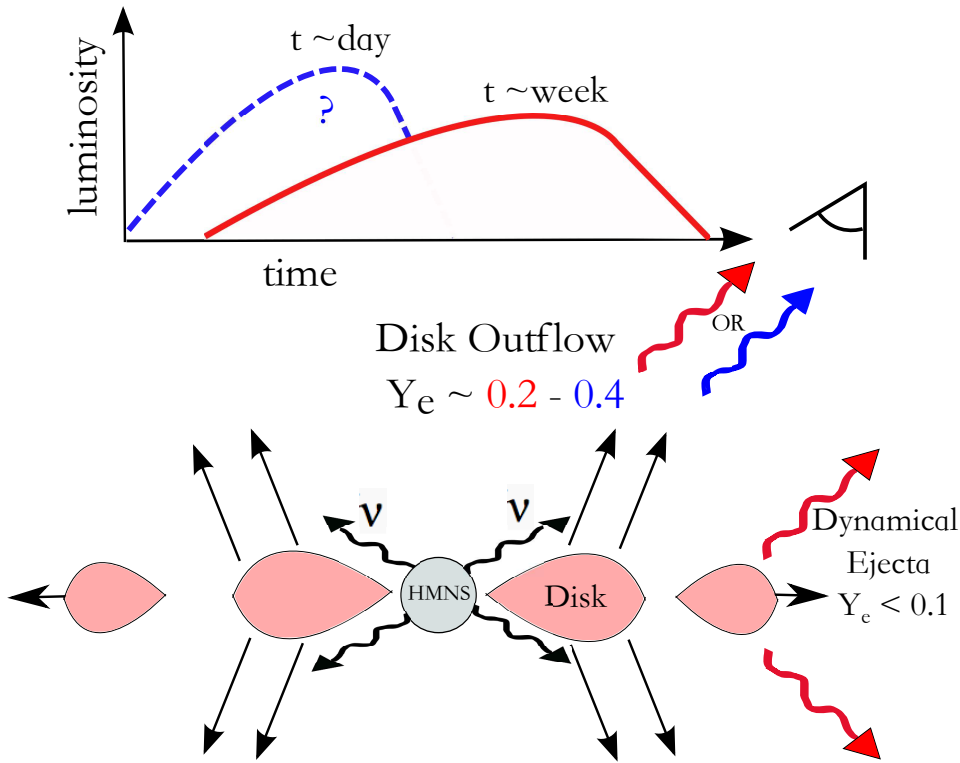}
\caption{
Relation between the observed kilonova and the properties
of the ejecta that powers it. Material ejected dynamically
in the equatorial plane is highly neutron rich ($Y_e < 0.1$),
producing heavy r-process elements that include Lanthanides. This results
in emission that peaks in the near-infrared and lasts for 
$\sim 1$~week (`late red bump') due to the high opacity.
Outflows from the remnant disk are more isotropic and also contribute
to the kilonova. If the HMNS is long-lived, then neutrino irradiation
can increase $Y_e$ to a high enough value ($Y_e\sim 0.4$) that no Lanthanides
are formed, resulting in emission peaking at optical wavelengths (`early blue bump').
If BH formation is prompt, outflows from the disk remain neutron
rich, and their contribution is qualitatively similar to that of the dynamical
ejecta.
}
\label{f:cartoon}
\end{figure*}

When a HMNS does form, its lifetime before collapsing into a BH depends on
the timescale for thermal energy loss via neutrino emission  
(e.g., \citealt{ruffert1999,Paschalidis+12,Galeazzi+13}) 
and the efficacy of angular momentum transport via gravitational
waves and magnetohydrodynamic stresses (e.g., \citealt{Duez+06,Stephens+08,Siegel+13}).
For particularly nearby mergers, 
oscillations excited in
the HMNS may be detectable in the GW strain data (\citealt{Shibata05};
\citealt{Bauswein+12}; \citealt{Hotokezaka+13c}).  However, the subsequent ring
down phase following BH formation is unlikely to be detected by the initial
generation of advanced detectors.  

Fortunately, NSMs are also accompanied by coincident electromagnetic (EM)
signals that inform physical processes at work during the merger
(e.g.~\citealt{Metzger&Berger12}; \citealt{Kelley+13}; \citealt{Piran+13}).
One such counterpart is a thermal IR/optical transient powered by the
radioactive decay of heavy elements synthesized in the merger ejecta (a
`kilonova'; \citealt{Li&Paczynski98}; \citealt{Metzger+10};
\citealt{Roberts+11}; \citealt{Goriely+11}; \citealt{Piran+13}; 
\citealt{Grossman+13}; \citealt{Tanaka+14}).  Kilonovae
are particularly promising EM counterparts because (1) their generation is
relatively robust, requiring only a modest amount of unbound ejecta; (2) their
signal is independent of the existence of a dense surrounding external medium;
and (3) unlike a GRB, kilonovae are relatively isotropic.  
A candidate kilonova was recently detected following the 
GRB 130603B (\citealt{Tanvir+13}; \citealt{Berger+13}; ).

If the merger ejecta is sufficiently neutron-rich
for $r$-process nucleosynthesis to
reach the Lanthanides ($A \gtrsim 139$), the optical opacity becomes much
higher than that of iron-group elements \citep{Kasen+13},
resulting in emission that is redder, dimmer, and more slowly evolving
(\citealt{Barnes&Kasen13,tanaka2013}).  Although such unusually red colors may be
beneficial in distinguishing NSM transients from unrelated astrophysical
sources, the current lack of sensitive wide field infrared telescopes could
make EM follow-up across the large sky error regions provided by Advanced
LIGO/Virgo even more challenging (e.g.~\citealt{Nissanke+13}; \citealt{Metzger+13}; \citealt{Hanna+13}; \citealt{Kasliwal&Nissanke13}).

The matter ejected dynamically following a NSM 
is likely to be sufficiently neutron-rich (as quantified by the electron 
fraction $Y_e~\lesssim 0.3$) 
to produce a red kilonova 
(e.g., \citealt{Rosswog05,Duez+09,bauswein2013}).
Dynamical expulsion is not the only source of ejecta, however.  A
robust consequence of the merger process is the formation of a remnant torus
surrounding the central HMNS.  Outflows from this accretion disk over longer,
viscous timescales also contribute to the merger ejecta
(e.g.,~\citealt{Surman+08,Metzger+08,Metzger+09a,Lee+09,Dessart+09,Wanajo&Janka12}).  
The more isotropic geometry of disk winds
suggests that 
they may contribute a distinct component to the
kilonova light curve for most viewing angles \citep{Barnes&Kasen13, Grossman+13}.

\citet[ hereafter FM13]{FM13} calculated the 
viscous evolution of remnant
BH accretion disks formed in NSMs using two-dimensional, time-dependent
hydrodynamical simulations.  Over several viscous times, FM13 found that 
a fraction $\sim$several percent
of the initial disk mass is ejected as a moderately
neutron-rich wind ($Y_e \sim 0.2$) powered by viscous heating and nuclear
recombination.  Although the higher entropy of the outflow as compared to the
dynamical ejecta results in subtle differences in composition (e.g. a small
quantity of helium), the disk outflows likely produce Lathanide elements with
sufficient abundance to result in a similarly 
red kilonova as with the dynamical ejecta.

FM13 included the effects of self-irradiation by neutrinos on the dynamics and
composition of the disk. Due to the relatively low
accretion rate and radiative efficiency at the time of the peak outflow, 
neutrino absorption had a sub-dominant contribution to the disk evolution.
This hierarchy
is important because a large neutrino flux tends to
drive $Y_e$ to a value 
higher than that in the disk midplane (e.g.~\citealt{Surman+08}; \citealt{Metzger+08}; \citealt{Surman+13}).  If
neutrino irradiation is sufficient to drive $Y_e \gtrsim 0.3-0.4$, 
the nuclear composition of the disk outflows would be significantly altered,
resulting in a distinct additional component visible in the kilonova emission. 

By ignoring the influence of a central HMNS, FM13 implicitly assumed a
scenario in which BH formation was prompt or the HMNS lifetime very short. 
Here we extend the study of FM13 
to include the effects of neutrino irradiation from a long-lived HMNS.  
As we will show, the much larger neutrino luminosity of the HMNS has 
a profound effect on the quantity and composition of the disk outflows, allowing
a direct imprint of the HMNS lifetime on the kilonova (Figure~\ref{f:cartoon}).
As in FM13, our study includes many approximations that enable
us to follow the secular evolution of the system. We focus here on exploring
the main differences introduced by the presence of a HMNS, and
leave more extensive parameter space studies or realistic computations
for future work.

The paper is organized as follows.  In $\S\ref{s:model}$ we describe the
numerical model employed. Our results are presented in $\S\ref{s:results}$, separated
into dynamics of the outflow (\S\ref{s:evolution}) and composition (\S\ref{s:composition}).  
A summary and discussion follows in $\S\ref{s:discussion}$.
Appendix~\ref{s:neutrino_details} describes in more
detail the upgrades to the neutrino physics implementation relative to that of FM13.

\section{Numerical Model}
\label{s:model}

Our numerical model largely follows that 
described in FM13. Here we summarize
the essential modifications needed to model
the presence of a HMNS.

\subsection{Equations and Numerical Method}
\label{s:numerical}

We use FLASH3.2 \citep{dubey2009} to solve 
the time-dependent hydrodynamic equations in two-dimensional, 
axisymmetric spherical geometry.
Source terms include the pseudo-Newtonian potential of \citet{paczynsky1980},
an anomalous shear stress for angular momentum transport
(\S\ref{s:angz_transport}), and charged-current
weak interaction terms in the energy and lepton number
equations (\S\ref{s:neutrinos}). The equation of state is that of 
\citet{timmes2000}, with the abundances of neutrons, protons,
and alpha particles satisfying nuclear statistical equilibrium, and
including the nuclear binding energy of alpha particles in the internal
energy\footnote{We have corrected an error in the treatment of the nuclear
binding energy of $\alpha$ particles, which led to an overestimation of the
recombination heating in the models of FM13. With the correct treatment, 
the amount of mass ejected decreases by a factor of $\sim$few and the ejection
occurs over a longer period of time relative to the values published in FM13.  
Changes to other properties 
relevant for nucleosynthesis ($Y_e$, entropy) are insignificant.}. 
The upper density limit in the tabulated
lepton component is extended using analytic expressions for a 
fully relativistic, arbitrary-degeneracy lepton gas (e.g., \citealt{bethe80}).

The code contains modifications relative to the public version, aimed at 
investigating the viscous evolution of merger remnant accretion 
disks (\citealt{F12,FM12}; FM13). 
The radial cell spacing is non-uniform, with consecutive cells having 
a constant ratio between their sizes. The meridional grid is 
uniform in $\cos\theta$. Models are initialized from an equilibrium 
torus that is allowed to relax for $100$ orbits -- without source terms -- 
to smooth out initial discontinuities.
This equilibrium initial solution adjusts its angular momentum profile to
near Keplerian on the local viscous time once
source terms are turned on. We
do not expect qualitative differences in our results being introduced by a different
initial angular momentum profile.

\subsection{Boundary Conditions and Angular Momentum Transport}
\label{s:angz_transport}

We approximate the HMNS by a reflecting inner boundary 
at a fixed radius $R_{\rm NS}= 30$~km
for all variables except specific angular momentum,
motivated by the approximate location of the neutrinosphere in the models of \citet{Dessart+09}. 
Our approximation is justified in that HMNSs from self-consistent calculations
achieve a state of quasi-equilibrium over a few dynamical times after the merger,
particularly if they are not too close to the
mass for prompt BH formation (e.g., \citealt{sekiguchi2011}). While our spherical
boundary does not initially follow the highly elliptical form of isobaric surfaces, the
accumulation of matter above the boundary develops such an elliptical shape after 
a few orbits of evolution. Whenever the NS is assumed to collapse to a BH, this reflecting
boundary condition is changed to absorbing for all variables as in FM13. The outer radial
boundary allows matter to leave the domain, while the symmetry axis is reflecting in the
meridional direction.

Angular momentum transport is mediated by an anomalous shear stress as in FM13.
The coefficient of kinematic viscosity is that of \citet{shakura1973}.
Whenever the HMNS is present, the specific angular momentum in the inner radial
ghost cells is set so that uniform rotation with an angular velocity $\Omega_\star
= 2\pi/P_\star$ is obtained (here $P_\star$ is the assumed rotation period of the
neutron star). The viscous stress is applied at the inner radial boundary, thus
driving the inner most active cells to co-rotate with the neutron star, and
resulting in the formation of a boundary layer (e.g.,
\citealt{frank2002}).\footnote{We neglect the spin-up (or spin-down) of the
neutron star by this torque, however, because the angular momentum contained in
the disk is small compared to that in the star.}  Although the MRI is unlikely
to operate in the boundary layer itself, waves produced by this interface can
result in angular momentum transport of a similar magnitude \citep{belyaev2013}.

The steep density gradient that develops in the boundary layer 
requires verifying that results are converged. We elaborate on 
this in Appendix~\ref{s:neutrino_details}.

\begin{table*}
\centering
\begin{minipage}{18cm}
\caption{Models Evolved and Outflow Properties\label{t:models}\label{t:results}. The first
six columns show model name, HMNS lifetime, stellar period, fractional radial resolution
at inner boundary, and irradiation by the neutron star, respectively. 
The last eight columns show integrated properties of the outflow, restricted to equatorial
 ($\theta \in [30^\circ,150^\circ]$) and polar ($\theta<30^\circ$ and $\theta > 150^\circ$)
latitudes. Each group of four columns includes the mass-flux weighted electron fraction, entropy,
and expansion time at a radius where the mass-flux weighted temperature is $5\times10^9$~K, as
well as the net mass in unbound material crossing a surface at $10^9$~cm, normalized 
by the initial torus mass.}
\begin{tabular}{lccccccccccccc}
\hline
{Model}&
{$t_{\rm ns}$} &
{$P_\star$} &
{$\alpha$} &
{$\Delta r_{\rm min}/R_\star$} &
{$\star$-Irr.} &
\multicolumn{4}{c}{Equatorial Outflow} &
\multicolumn{4}{c}{Polar Outflow$^b$}\\
{ } & \multicolumn{2}{c}{(ms)} &  {}  &  {}  & {} & 
{$\bar{Y}_e$} &
{$\bar{s}$ } &
{$t_{\rm exp}$} &
{$M_{\rm ej,unb}/M_{\rm t0}$} &
{$\bar{Y}_e$} &
{$\bar{s}$ } &
{$t_{\rm exp}$} &
{$M_{\rm ej,unb}/M_{\rm t0}$}\\
\multicolumn{6}{c}{} & {} & {($k$/b)} & {(ms)} & {(\%)} & {} & {($k$/b)} & {(ms)} & {(\%)}\\
\hline
t000A3      & 0        & ... & 0.03 & 5E-3 & No   & 0.18 & 18 & 27  & 2.1 & ...  & ... & ... & 0.9 \\
t010A3p15   & 10       & 1.5 &      &      & Yes  & 0.19 & 20 & 30  & 2.5 & ...  & ... & ... & 1.0 \\
t030A3p15   & 30       &     &      &      &      & 0.23 & 21 & 27  & 3.9 & 0.50 & 45  & 7.6 & 1.3 \\
t100A3p15   & 100      &     &      &      &      & 0.28 & 20 & 33  & 9.2 & 0.47 & 42  & 11  & 4.0 \\
t300A3p15   & 300      &     &      &      &      & 0.30 & 20 & 32  & 28  & 0.44 & 38  & 17  & 8.8 \\
tInfA3p15   & $\infty$ &     &      &      &      & 0.32 & 21 & 42  & 67  & 0.43 & 52  & 17  & 22  \\
\noalign{\smallskip}
pA1p15      & $\infty$ & 1.5 & 0.01 & 5E-3 & Yes  & 0.36 & 21 & 61  & 47  & 0.45 & 52  & 19  & 22 \\
pA3p20      &          & 2   & 0.03 &      &      & 0.30 & 29 & 37  & 62  & 0.37 & 62  & 15  & 17 \\
pA3p15xs    &          & 1.5 &      &      & No   & 0.27 & 21 & 45  & 56  & 0.35 & 61  & 15  & 23 \\
\noalign{\smallskip}
rA3p15r1    & $\infty$ & 1.5 & 0.03 & 1E-2 & Yes  & 0.32 & 21 & 45  & 69  & 0.43 & 53  & 17  & 22 \\
rA3p15r2$^a$   &       &     &      & 2E-3 &      & ... & ... & ... & ... & ...  & ... & ... & ... \\
\hline
\hline
\multicolumn{14}{l}{$^a$ Model rA3p15r2 was evolved for a shorter time. See Appendix~\ref{s:neutrino_details} 
	            for details.}\\
\multicolumn{14}{l}{$^b$ Average thermodynamic quantities are not computed if there is no region
			 where the average temperature is $\sim 5\times 10^9$~K.}
\label{table:models}
\end{tabular}
\end{minipage}
\end{table*}

\subsection{Neutrino Treatment}
\label{s:neutrinos}

We consider contributions to the neutrino irradiation from both the central
HMNS and from the disk.  The neutrino flux from the HMNS is assumed to be
spherically symmetric and isotropic, following a Fermi-Dirac spectrum with zero
chemical potential and a constant neutrinospheric temperature. 
We take equal luminosities of electron neutrinos and antineutrinos, with
temperatures $T_{\nu_e,{\rm ns}}=4$~MeV and $T_{\bar{\nu}_e,{\rm ns}}=5$,
respectively, yielding mean neutrino energies of $\sim 12$ and $15$~MeV,
respectively.
We ignore the highly aspherical form of the neutrino flux and temperature obtained in
simulations with more sophisticated neutrino transport (e.g., \citealt{Dessart+09}).
Since the viscously-driven winds arise on timescales of a few seconds, we must
account for the change in the neutrino luminosity due to the cooling evolution
of the HMNS (e.g., \citealt{Roberts12}).  We do this by adopting a temporal fit to
the cooling curves of \citet{pons1999}, with a normalization that approximately
matches the luminosities at $30$~ms obtained by \citet{Dessart+09}, $2\times 10^{52}$~erg~s$^{-1}$.  
The neutrino flux incident upon a given point in the computational domain is
attenuated by the integrated optical depth from the inner boundary
along radial rays.  

Given the large optical depths obtained in the boundary layer,
we replace the neutrino cooling implementation used in FM13 -- which was appropriate
for a disk of low to moderate optical depth -- with 
a neutrino leakage scheme that follows that of \citet{ruffert1996}. 
The spatial emission properties of the disk itself are also slightly more complicated
when a HMNS sits at the center than when a black hole is the central object, thus
our prescription for disk \emph{self-irradiation} must be modified. A detailed
description of these changes is documented in Appendix~\ref{s:neutrino_details}. 

Charged-current weak interaction rates are otherwise unmodified relative to
FM13. Additional neutrino emission channels are sub-dominant and hence are
neglected. We impose a floor on the electron fraction at $Y_{e,{\rm
min}} = 0.01$ to prevent problems with our tabulated rates. Finally, we neglect
the neutrino contributions to the pressure and internal energy in optically
thick regions: this contribution is at most $\sim 10\%$ in the equatorial part
of the boundary layer.

\subsection{Models Evolved}

The models run are summarized in Table \ref{table:models}.  Our baseline
parameter set consists of a central object (BH or HMNS) with mass $3M_\sun$, a
disk mass $M_{\rm t0} = 0.03M_\sun$, constant specific angular momentum, and a
`disk radius' $R_0 = 50$~km. The entropy and electron fraction are uniform,
initially set to $8k_{\rm B}$ per baryon and 0.1, respectively. We use $384$
cells in radius, with an innermost radial spacing set to $\delta r_{\rm
min}/R_\star\simeq 5\times 10^{-3}$ in the fiducial set of models. A total of
112 cells are used in the $\theta$ direction. 
Most models are evolved for $3000$ orbits at $r=R_0$, or $\sim 8.7$~s.  

Six models probe the effect of the HMNS lifetime $t_{\rm ns}$ on the mechanical
and compositional properties of the disk wind (t-series). The model with zero
HMNS lifetime (t000A3, pure BH) has the same set of parameters as the fiducial model
used in FM13, but differs in its neutrino implementation
(\S\ref{s:neutrinos}) and in the treatment of the nuclear binding energy
of $\alpha$ particles (\S\ref{s:numerical}). 

Another set of three models (p-series) probes the effect of varying important
parameters relative to the baseline model with an infinite HMNS lifetime
(tInfA3p15).  We explore the sensitivity to the assumed rotation period
$P_\star$, the magnitude of the viscous stress $\alpha$, and the effect of
removing neutrino irradiation by the HMNS.
Finally, two models (r-series) quantify the sensitivity of our results on the
radial resolution employed near the inner boundary 
(Appendix~\ref{s:neutrino_details}).

\begin{figure*}
\includegraphics*[width=\textwidth]{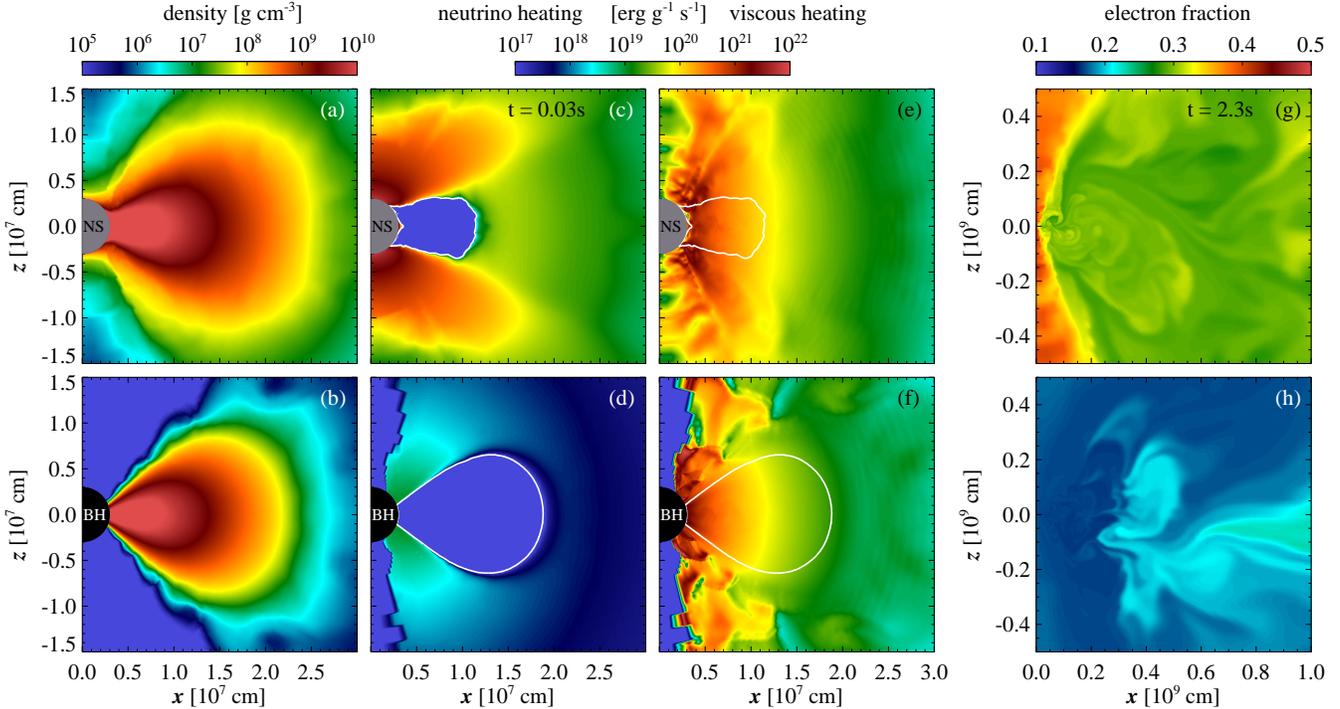}
\caption{Snapshots in the evolution of models tInfA3p15 (stable HMNS, top)
and t000A3 (prompt BH, bottom). The nine contiguous panels on the left show density (a,b), neutrino
heating (c,d), and viscous heating (e,f) at time $0.03$s, corresponding approximately
to $10$ orbits at the initial density peak. The two panels on the right (g,h) show
the electron fraction at time $2.3$s (800 orbits at the initial density peak). The
white contours show the gain surface, inside which emission of neutrinos and
antineutrinos dominates over their absorption by nuclei. Models suppress
neutrino and viscous source terms below a density of $10$~g~cm$^{-3}$ (FM13) for numerical reasons 
(as in the polar region of panels d and f).}
\label{f:2d_map}
\end{figure*}

\vspace{0.9in}

\section{Results}
\label{s:results}

\subsection{Evolution and Mass Ejection}
\label{s:evolution}

The presence of a HMNS results in two important differences
relative to the case where a BH forms promptly.
To illustrate these differences, we first focus the 
discussion on models tInfA3p15 and t000A3, corresponding
to infinitely lived HMNS and prompt BH, respectively.

First, a hard stellar surface leads to the formation of
a boundary layer over the first few orbits as the disk begins 
to accrete. Material that has angular momentum removed by
the viscous stress migrates towards the polar regions of the stellar
surface, forming an envelope (Figure~\ref{f:2d_map}a). 
In contrast, a BH accretion disk is such that the density 
undergoes a steep decline at an angle from the midplane, with
the polar regions being largely devoid of significant material (Figure~\ref{f:2d_map}b).

The second important difference introduced by the HMNS is
the level of neutrino heating in the system. The geometry of
this heating is shown in Figure~\ref{f:2d_map}c. Irradiation
by the HMNS is strongest in the polar regions, where 
there is less attenuating material. Emission in this
region is also contributed to by the disk itself, which emits mostly
from two emission spots above the midplane near the stellar 
surface.
While regions near the disk midplane are dominated by neutrino cooling
at inner radii, neutrino heating dominates cooling outside $\sim 100$~km 
even when the irradiation from the star is shadowed by the disk.
These emission properties stand
in stark contrast to the case of a BH (Figure~\ref{f:2d_map}d), in which
neutrino heating is largely confined to the polar regions, and at a
much lower level, which makes it insignificant relative to viscous heating (FM13).

\begin{figure}
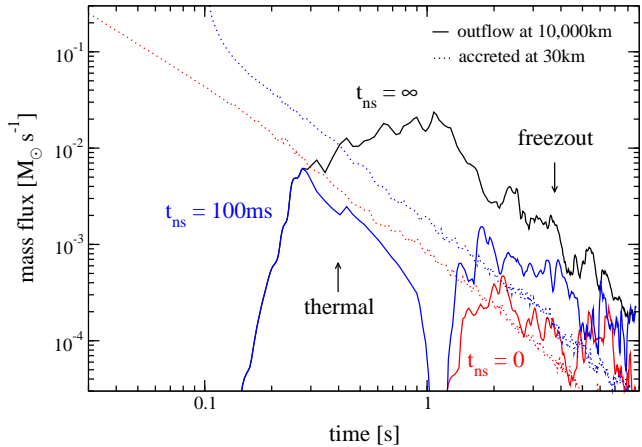

\begin{overpic}[width=\columnwidth]{f3.eps}
\put(3.5,45.5){\tiny $\odot$}
\end{overpic}
\caption{Mass loss rate in unbound material at 10,000~km (solid lines)
and net accretion rate at $30$~km (dotted lines) as a function of time.
Shown are three models that illustrate the difference between
prompt BH formation (t000A3p15, red), infinitely 
lived HMNS (tInfA3p15, black), and an intermediate case
with $t_{\rm ns}=100$~ms (t100A3p15). Mass fluxes are computed
over the full range of polar angles. Mass ejection before
$1$~s occurs on the thermal time of the disk, while the late-time wind
happens due to weak freezout \citep{metzger09b}.}
\label{f:mass_loss}
\end{figure}

The combined action of these two effects -- boundary layer and 
higher degree of neutrino heating --
leads to very different mass ejection properties depending on the
nature of the central object (Figure~\ref{f:mass_loss}).
On the one hand, the mass ejection mechanism from a BH accretion disk 
relies on the \emph{weak freezout} of the disk
(e.g., \citealt{metzger09b}), in which neutrino cooling
shuts down as the disk spreads viscously and the temperature decreases.
Viscous heating, nuclear recombination, and transport of rotational
kinetic energy outwards all act as positive energy source terms which
are uncompensated at radii $\gtrsim 100$~km, leading to ejection of the
outer layers of the disk. This ejection manifests as a broad equatorial 
outflow ($\sim 60^\circ$ from the equator) that carries
several percent the initial disk mass on a timescale of a few seconds (FM13).

In contrast, a HMNS disk 
displays a distinct phase of mass ejection that operates on the thermal time of
the outer disk ($\sim 30-100$~ms for $r\sim 100-200$~km) due to the larger
amount of neutrino heating
and the concentration of neutrino cooling in a smaller spatial 
region (Fig.~\ref{f:2d_map}c). This phase occurs
well before weak freezout.  The presence of a reflecting boundary condition also
implies that inward traveling sound waves -- generated when material 
is heated and expands -- are available for pressure buildup instead
of being absorbed by a BH. The \emph{equatorial outflow} is thus significantly
enhanced when the HMNS is present for a time much longer than the 
thermal time of the disk.

In addition, strong neutrino heating at polar latitudes in HMNS disks
causes a genuine \emph{neutrino-driven wind} at early times in 
this direction (see, e.g., \citealt{Dessart+09}).
After several $10$~ms, viscous heating takes over neutrino heating.
Overall, mass loss along the polar directions goes from less than $1\%$ of the disk 
mass in the pure BH case to nearly $\sim 20\%$ for an 
infinitely lived HMNS (Table~\ref{t:models}).
Given that this material 
is strongly irradiated, its composition is qualitatively different from
that ejected along the disk midplane (\S\ref{s:composition}). 

\begin{figure*}
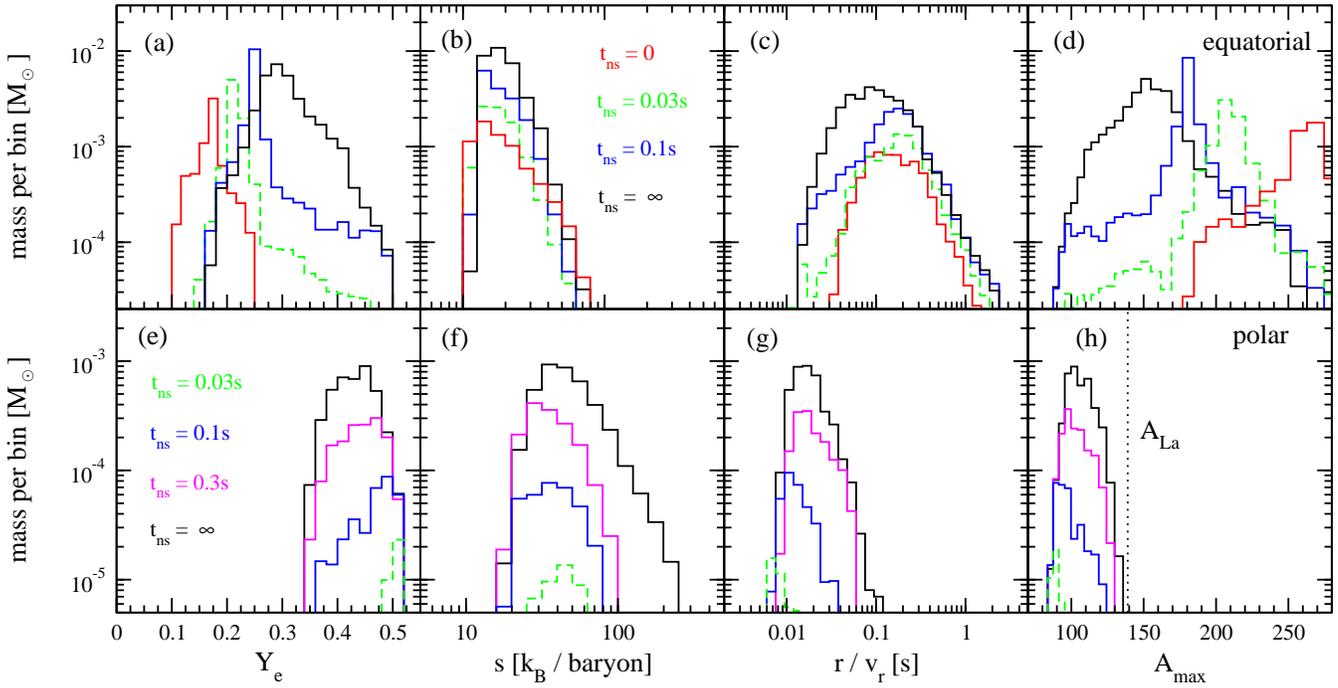

\begin{overpic}[width=\textwidth]{f4.eps}
\put(1.3,23){\tiny $\odot$}
\put(1.3,45.65){\tiny $\odot$}
\end{overpic}
\caption{Mass histograms of material with positive velocity, evaluated
at radii such that the mass-flux weighted temperature is $\sim 5\times 10^9$~K.
The counting is restricted to latitudes within $60^\circ$ of the midplane
(top) and within $30^\circ$ of the polar axis (bottom).
Shown are the distributions of electron fraction (a,e), entropy (b,f),
expansion time (c,g), and maximum mass number of r-process
elements (d,h) obtained from equation (\ref{eq:Amax}), 
assuming seed nuclei with $\bar A = 90$ and $\bar Z = 36$. Different curves correspond
to models with different HMNS lifetime, as labeled. The polar outflow
has a peak $A_{\rm max}$ below the Lanthanides ($A \gtrsim A_{\rm La} = 139$).}
\label{f:histogram}
\end{figure*}

Allowing the HMNS to survive for different lengths of time results in a behavior
that is intermediate between the pure BH and infinitely lived HMNS, as shown
in Figure~\ref{f:mass_loss} for a HMNS with a lifetime of $100$~ms. The 
ejected mass depends monotonically on the HMNS lifetime 
(Table~\ref{t:models}),
at both polar and equatorial latitudes. Polar outflows are significant if
$t_{\rm ns}\gtrsim 100$~ms.

The dynamics of the flow at the transition from HMNS to BH are characterized
by instantaneous accretion of the boundary layer material, and subsequent
generation of a rarefaction wave. 
The velocity becomes negative from inside out, and the initial phase of accretion is
interrupted. This cutoff in the mass loss is quite steep, as shown
in Figure~\ref{f:mass_loss} for the model with $t_{\rm ns} = 100$~ms.
The equatorial regions of the disk relax to the BH configuration, 
and the late-time outflow appears at the expected
time of $\sim 1$~s. The fluid at polar latitudes, which was previously supported
mostly by gas pressure, collapses into the hole leaving an evacuated
polar cavity. A fraction of the material still manages to escape, however, as
shown in Table~\ref{t:models}.

Decreasing the viscosity of the disk (model pA1p15) relative
to the fiducial $t_{\rm ns}=\infty$ case (tInfA3p15) results in a 
smaller amount of equatorially ejected material. The main
reason for this difference lies in that viscous heating is
more centrally concentrated for lower $\alpha$. 
The accretion luminosity is also smaller by a factor of $\sim 2$, 
resulting in less neutrino heating in equatorial regions on the
thermal timescale of the disk. The time-integrated neutrino energy
deposition at polar latitudes differs only by $\sim 10\%$, however.

Increasing the rotation period of the star (model pA3p20) results in less
mass ejected along the poles, with a smaller decrease in the equatorial
ejection relative to the fiducial model. This result can be traced back to the behavior of the
boundary layer. Since the fluid is forced to co-rotate with the star
at the surface, a lower rotation rate implies a lower centrifugal 
force, which allows the boundary layer to spread more material
to high latitudes (the component of the centrifugal force tangential
to the stellar surface points towards the midplane; e.g., \citealt{inogamov2010}).
This excess of material is able to attenuate more neutrino flux,
decreases the heating, resulting in less mass ejection (with
a smaller electron fraction).

Finally, removing irradiation from the HMNS while keeping the reflecting
boundary and self-irradiation from the disk (model pA3p15xs) results
in a lower amount of mass ejected in equatorial regions.
As in the case of lower $\alpha$, there is less neutrino heating
at larger radii on the disk midplane. The decrease in the equatorial
mass ejection relative to the fiducial model is less than in 
the low-$\alpha$ case, however, because viscous heating is unchanged.
The fact that the amount of mass ejected along the poles is
comparable (with smaller $Y_e$, however) shows that viscous
heating is also a fundamental agent in driving the polar outflow.

\subsection{Composition}
\label{s:composition}

Free nuclei recombine into $\alpha$-particles once the temperature decreases to
$T \lesssim 10^{10}$ K.  Heavier elements start to form once the temperature
decreases further, $T \lesssim 5\times 10^{9}$, via the reaction
$^{4}$He($\alpha$n,$\gamma$)$^{9}$Be($\alpha$,n)$^{12}$C.  After $^{12}$C
forms, additional $\alpha-$captures produce heavy `seed' nuclei with
characteristic mass $\bar{A} \simeq 90-120$ and charge $\bar{Z} \simeq 35$ (the
`$\alpha$-process'; \citealt{Woosley&Hoffman92}).  Whether nucleosynthesis
proceeds to heavier $r$-process nuclei of mass $A$ depends on the ratio of free
neutrons to seed nuclei once the $\alpha$-process completes.  Since the
formation of $^{12}$C is the rate-limiting step in forming seeds, this critical
ratio  depends primarily on three quantities (e.g.~\citealt{Hoffman+97};
Appendix C of FM13): the electron fraction $Y_e$, entropy $S$, and the
expansion timescale $t_{\rm exp}$ at times just following $\alpha$ particle formation
($T \sim 5\times 10^{9}$ K).  

A mass-flux average of each of these three quantities is given in 
Table \ref{table:models} for most models (see FM13 for a description
of the calculation method). The radial position for the average is chosen
so that the mass-flux averaged temperature is approximately $5\times 10^9$~K. 
The average is separated between
equatorial and polar latitudes ($60^\circ$ from the midplane and $30^\circ$
from the axis, respectively). In the case of prompt BH formation or very
shortly lived HMNS ($t_{\rm ns} \leq 10$~ms), there is never enough material
in the polar regions to achieve the desired temperature, and hence
the average is not computed there. After collapse to a BH, the polar
region is evacuated, so the radius for the average is obtained for
times less than the HMNS lifetime (but averaged thermodynamic
quantities are computed using the whole evolution).

The average electron fraction of the material is a monotonic function
of the HMNS lifetime. This is a direct consequence of the higher
level of neutrino irradiation introduced by the HMNS. A rough estimate
of the change in $Y_e$ over a thermal time in the boundary layer yields
\begin{eqnarray}
\Delta Y_e & \sim & \frac{Q_\nu \Delta t}{\langle \varepsilon_\nu\rangle/m_n}\nonumber\\
&\sim & 1\left(\frac{Q_\nu}{10^{21}\textrm{ erg g}^{-1}\textrm{ s}^{-1}} \right)
\left( \frac{10\textrm{ MeV}}{\langle\varepsilon_\nu\rangle}\right)
\left(\frac{\Delta t}{10\textrm{ ms}}\right),
\end{eqnarray}
where $Q_\nu$ is the specific neutrino heating rate (Figure~\ref{f:2d_map}c),
$\langle\varepsilon_\nu\rangle$ is the mean neutrino energy, and $\Delta t$
is the time interval. Irradiation by the HMNS can thus introduce changes of order
unity in the electron fraction over the time it takes the polar
outflow to be launched. Similar considerations apply to the equatorial
outflow, although the fact that this ejecta originates in regions
of the disk that are both farther out in radius and shadowed by the
inner regions causes the changes in $Y_e$ to be less pronounced.

It is worth keeping in mind that we are imposing equal luminosities
of electron neutrinos and antineutrinos from the HMNS (\S\ref{s:neutrinos}). In the 
limit of high irradiation, neutrinos drive $Y_e$ towards \citep{Qian&Woosley96}
\begin{equation}
Y_e^{\rm eq} \simeq \left[1 + \frac{\epsilon_{\bar{\nu}_e}-\Delta+\Delta^2/\epsilon_{\bar\nu_e}}
			           {\epsilon_{\nu_e}      +\Delta+\Delta^2/\epsilon_{\nu_e}}\right]^{-1},
\end{equation}
where $\Delta = 1.293$~MeV is the neutron-proton mass difference, and $\epsilon_{\nu_i}$
is the ratio of the mean square energy to the mean energy of the 
distribution, $\langle E_\nu^2\rangle/\langle E_\nu\rangle$. Our Fermi-Dirac spectrum with 
zero chemical potential implies 
$\epsilon_{\nu_i}\simeq 4kT_{\nu_i}$, hence $Y_e^{\rm eq} \simeq 0.52$.

\begin{figure}
\includegraphics*[width=\columnwidth]{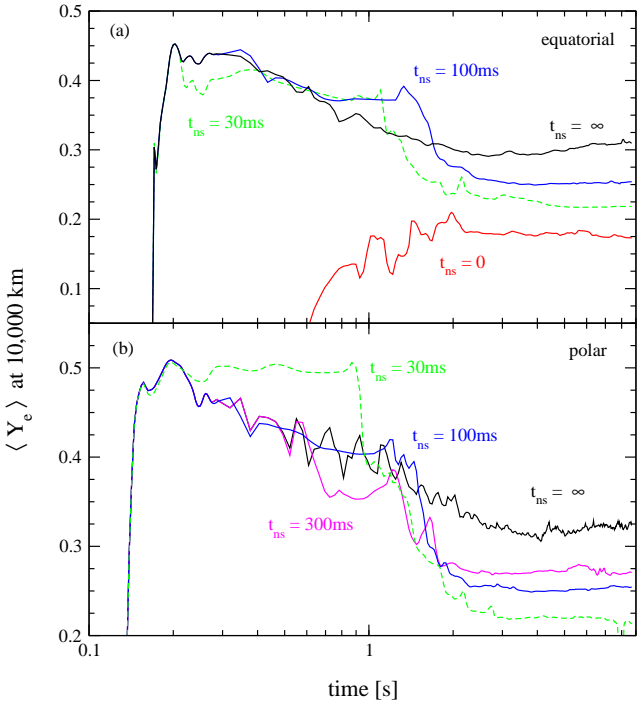}
\caption{Mass-flux weighted electron fraction at $10,000$~km 
as a function of time, for models with different HMNS lifetime $t_{\rm ns}$.
The averages are restricted to latitudes within $60^\circ$ of the midplane
(top) and within $30^\circ$ of the polar axis (bottom). The drop around $1-2$~s
for the polar material arises from the combination of a decreasing 
neutrino luminosity and the travel time to $10,000$~km.
Note also that the vertical scales are different.}
\label{f:ye_average}
\end{figure}

The distribution of thermodynamic properties
of material with positive velocity -- evaluated at the same radii as
the averages in Table~\ref{t:models} -- is shown in Figure~\ref{f:histogram}.
The electron fraction of the equatorial material has a peak that
tracks the average value, with a long tail to high $Y_e$
for longer HMNS lifetime. In contrast, the polar material has more
material with lower $Y_e$ for longer $t_{\rm ns}$. This is a reflection
of the time-dependence of the outflow composition, which is
dominated by neutron-rich conditions at late time (Figure~\ref{f:ye_average}).

The distribution of entropy in equatorial material is insensitive to
the lifetime of the HMNS. The polar material also has its peak nearly
unchanged, though a high-entropy tail develops for longer $t_{\rm ns}$. 
A similar behavior is seen in the expansion time, with a peak values that
undergo moderate shifts for both polar and equatorial outflows, but with
changes in the shape of the distribution for longer-lived HMNSs.

The maximum mass number $A_{\rm max}$ to which the $r$-process may proceed can 
be estimated as (\citealt{Hoffman+97})
\be
A_{\rm max} = \bar{A}\left[1 + \frac{X_{\rm n}}{1-X_{\rm n}-X_{\rm He}}\right],
\label{eq:Amax}
\ee
where $X_{\rm n}$ and $X_{\rm He}$ are the mass fractions of neutrons and alpha
particles, respectively, following the $\alpha$-process. 
We take the residual alpha fraction from the high-density trajectories
of \citet{Woosley&Hoffman92}, which have an expansion time $\sim 0.1$~s
and entropy $\sim 20$~k$_{\rm B}$ per baryon. Below the lowest electron
fraction tabulated ($Y_e = 0.425$), we assume that no alpha particles
form. The electron fraction is set to $X_n = 1 - 2Y_e$.

The resulting distribution of $A_{\rm max}$ (eq.~[\ref{eq:Amax}]) for
equatorial and polar outflows, obtained assuming $\bar A = 90$ and $\bar Z =
35$, is shown
in Figure~\ref{f:histogram}. 
The equatorial wind is expected to
make heavy r-process elements, with the peak $A_{\rm max}$ decreasing with
longer HMNS lifetime down to $\sim 150$ for $t_{\rm ns}\to \infty$. The
distribution is broad, however, with a characteristic width $\Delta A_{\rm
max}\sim 50$.  Note that a value of $A_{\rm max} \gg 200$ does not necessarily
imply that such heavy nuclei will form, but rather that some fraction will
fission to lighter nuclei $A \sim 130$ before capturing additional neutrons.

In contrast, the polar outflow maintains the composition of the seed
nuclei, with a minor contribution at higher mass number in the case
of a long-lived HMNS.
As shown in Figure~\ref{f:ye_average}, in all cases the late-time value of $Y_e$ is lower
than that in the initial episode of mass ejection. Hence the  tail in $A_{\rm
max}$ corresponds to material ejected at late times.  Even if the seed nuclei
are as heavy as $\bar A = 120$, the early episode of mass ejection will have
its peak below the Lanthanides.


\section{Summary and Discussion}
\label{s:conclusions}
\label{s:discussion}

We have explored the effects of a hypermassive neutron star (HMNS)
on the long-term evolution of remnant accretion disks formed
in neutron star binary mergers. Our main results can be summarized as follows:
\newline

\noindent
1. -- A long-lived HMNS results in the ejection of a significant
      fraction of the disk over a timescale of $\sim 1$~s. The
      amount of mass increases monotonically with HMNS lifetime.
      This enhanced mass loss, up to a factor $\gtrsim 10$ relative to prompt BH formation, 
      results from enhanced neutrino heating and a reflecting inner boundary.
      \newline

\noindent
2. -- The composition of the ejecta is latitude-dependent. 
      Material within an angle $\sim 30^\circ$ of the polar axis
      is strongly irradiated by the HMNS. If the neutrino and antineutrino luminosities are similar (as we have assumed), the electron fraction of the outflow can be raised to values where Lanthanides are not produced. 
      \newline

\noindent
3. -- Material ejected equatorially is still expected to produce
      a strong r-process, similar to that of the dynamical ejecta, although the detailed composition is
      dependent on the HMNS lifetime.
      \newline 

The criterion on the neutron star lifetime to appreciably change the electron
fraction of the polar outflow ($t_{\rm ns} \gtrsim 100$ ms) is obviously
satisfied if the remnant mass is below the maximum mass of a cold neutron star,
or if the remnant is
stabilized by solid body rotation (a {\it supramassive} NS), the latter of which
is only removed on much longer timescales via e.g. magnetic dipole spin-down.
The $t_{\rm ns} \gtrsim 100$ ms condition is also likely satisfied if the HMNS
is supported by thermal pressure (e.g.~\citealt{Bauswein+10};
\citealt{Paschalidis+12}; \citealt{kaplan2013}), as the latter is removed on the
HMNS cooling timescale, which is typically on the order of $\sim$ seconds
(e.g.~\citealt{pons1999}).  On the other hand, support via differential rotation
may not last 100 ms, as the latter can be efficiently removed by magnetic
stresses: the growth rate of the MRI is as short as milliseconds.

Whether a blue bump indeed develops out of the polar outflow in the case of a
long-lived HMNS will depend on a number of factors. First, enough mass needs to
be ejected so that the contribution to the lightcurve from radioactive decay
becomes detectable.  Second, the level of irradiation and the ratio between
neutrino and antineutrino luminosities must be sufficient to raise $Y_e$ to
values close to $0.5$.  The results of \citet{Dessart+09}, who employ a much
more realistic treatment of neutrino transport but did not include the viscous
evolution, seem to align with our findings. 
The ejecta mass required for a detectable signal depends on how much
radioactive heating is supplied by synthesized elements, which are lighter than
the $r$-process nuclei but potentially more neutron-rich than $^{56}$Ni
(e.g.~\citealt{Grossman+13}).  

A final requirement for producing a blue bump is that the high $Y_e$
(Lanthanide-free) material must be ejected first, such that it resides {\it
exterior} to any lower $Y_e$ (Lanthanide-rich) material that would otherwise
block its emission.  Our calculations support this requirement as well: after
rising quickly, $Y_e$ of the polar ejecta decreases monotonically with time
(Fig.~\ref{f:ye_average}). This is a direct consequence of the decrease in the
neutrino luminosities with time, either
smoothly as in the case of a long-lived HMNS 
(Appendix~\ref{s:neutrino_details}), or via a sudden drop when a BH forms.
Future work will explore in more detail the
observational consequences of this bimodal outflow.

\begin{figure}
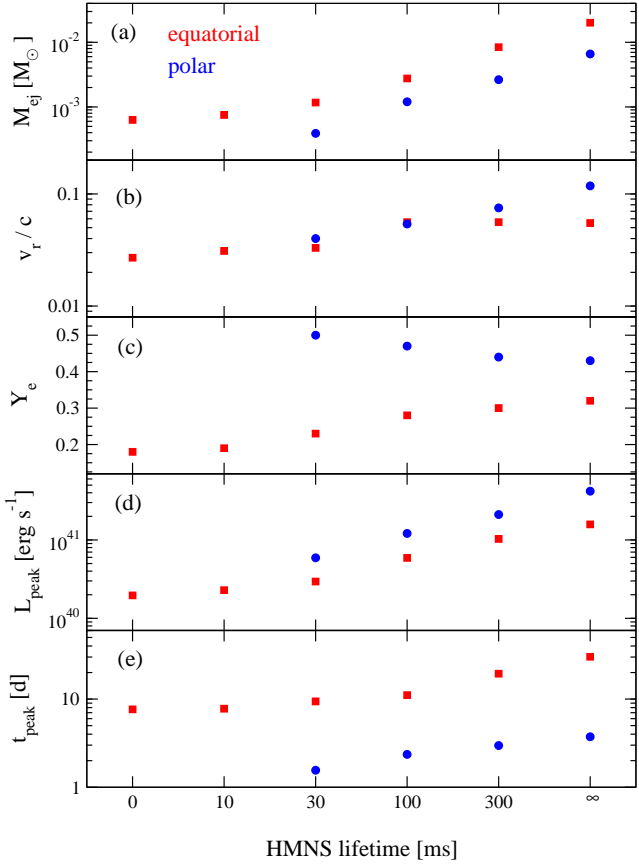

\begin{overpic}[width=\columnwidth]{f6.eps}
\put(2.5,93.3){\tiny $\odot$}
\end{overpic}
\caption{Kilonova properties as a function of HMNS lifetime. Shown are
the ejecta mass in unbound material (a), the mass-flux weighted velocity
at $10^9$~cm (b), the mass-flux weighted electron fraction (c), the
peak luminosity (d) (eq.~[\ref{eq:L_peak}]), and the time to peak (e) (eq.~[\ref{eq:t_peak}]).
Red squares and blue dots indicate equatorial and polar material. The opacity
is assumed to be Lanthanide-dominated for the former and Lanthanide-free for the latter.
Note that these numbers underestimate the `red' component since we have not
included contributions from the dynamical ejecta.
}
\label{f:summary_plot}
\end{figure}

Figure~\ref{f:summary_plot} provides estimates of various quantities relevant to the
kilonova emission derived from our calculations, as a function of 
the HMNS collapse time.  The peak luminosities and time to peak 
are estimated using the `Arnett rule' for a kilonova 
(e.g., \citealt{Li&Paczynski98,Metzger+10}),
\begin{eqnarray}
\label{eq:L_peak}
L_{\rm peak} & \simeq & 4.3\times 10^{41}\textrm{ erg s}^{-1}\left(\frac{f}{3\times 10^{-6}} \right)
		                     \left(\frac{v_r}{0.1c} \right)^{1/2}\nonumber\\
             &&\qquad\qquad\qquad\qquad \times \left(\frac{M_{\rm ej}}{0.01M_\sun} \right)^{1/2}\kappa^{-1/2}\\
\label{eq:t_peak}
t_{\rm peak} & \simeq & 1.4\textrm{ d } \left(\frac{v_r}{0.1c} \right)^{-1/2}
		\left(\frac{M_{\rm ej}}{0.01M_\sun} \right)^{1/2}\kappa^{1/2},
\end{eqnarray}
where $f$ is a factor quantifying radioactive energy deposition, and $\kappa$ is the opacity
of the material in units of cm$^2$~g$^{-1}$. In making Figure~\ref{f:summary_plot},
we have used $\kappa = 10$~cm$^2$~g$^{-1}$ for the equatorial material (Lanthanide-dominated),
and $\kappa = 1$~cm$^2$~g$^{-1}$ for the polar material (Lanthanide-free), see \citet{Kasen+13}.
The velocity is the mass-flux weighted value at $r = 10^9$~cm
considering only unbound material.

Most of the resulting kilonova properties are a monotonic function of the HMNS lifetime. Peak luminosities
increase from $\sim 10^{40}$ to $\sim 10^{41}$~erg~s$^{-1}$ for the red component, while the
blue component, when present, is brighter by a factor $\sim 2$. Similarly, the peak
times for the red component range from about a week to a month, increasing with longer 
HMNS lifetime due to the larger ejected mass. The blue component can last from a few- to 
several days. While the blue component is faster with longer HMNS lifetime, 
the average velocity of the red component saturates at $\sim 0.05c$.

The large amount of ejecta mass found by our calculations suggests that
outflows from the disk could easily overwhelm that from the dynamical ejecta.
One implication of this result relates to the Galactic production of
$r$-process elements.  FM13 estimated that ejection of $\sim 10\%$ of the disk
mass would contribute with $\sim 20\%$ of the production rate of elements with
$A\gtrsim 130$ assuming reasonable values for the disk mass and neutron star
merger rate. The increase in the ejected fraction of the disk by a factor of
several  relative to the prompt BH case implies that disks with long-lived HMNS
could become a dominant contribution to the galactic $r$-process element
production (\citealt{Freiburghaus+99}; \citealt{Rosswog+13};
\citealt{Piran+14}). This is the case even when the polar outflow is Lanthanide
free (Fig~\ref{f:histogram}).  

The large ejecta masses we infer for a moderately long-lived HMNS ($t_{\rm ns}
= 100$ ms) may also help alleviate the tension between large ejecta mass $\sim
3\times 10^{-2}M_{\odot}$ required to fit the NIR excess observed following GRB
130603B with current models (\citealt{tanaka2013}; \citealt{Piran+14}), without
the need to invoke less likely scenarios such as the merger of a NS with a low
mass BH.  If the long-lived HMNS is magnetized, its rotational spin-down could
also power the excess X-rays observed following this event (\citealt{Fong+14};
\citealt{Metzger&Piro13}; \citealt{Fan+13}).  Given that early blue emission
appears to require the presence of a long-lived HMNS, detection of such a
component in future events provides a relatively clean way to rule out a NS-BH
merger (although the {\it absence} of early blue emission would not rule out a
NS-NS merger).  

Our models include many approximations in order to make the evolution to a time
$\sim 10$ s computationally feasible.  In addition, we have focused here on the
key differences introduced by the HMNS applied to models with a particular
choice of parameters. Much more work remains in order to make reliable
predictions for kilonovae emission. In addition to a more extensive exploration
of parameter space,  models with realistic angular momentum -  and neutrino
transport will be needed. 


\section*{Acknowledgments}

We thank Dan Kasen, Eliot Quataert, Edo Berger, Gabriel Mart\'inez, Sasha Tchekhovskoy,
Wen-fai Fong, and Mansi Kasliwal for useful discussions and/or comments on the manuscript. 
We also thank the anonymous referee for helpful comments that improved the manuscript.
BDM acknowledges support from Columbia University.
RF acknowledges support from the University of California Office of the President, and
from NSF grant AST-1206097.
The software used in this work was in part developed by the DOE NNSA-ASC OASCR Flash Center at the
University of Chicago. 
This research used resources of the National Energy Research Scientific Computing
Center (NERSC), which is supported by the Office of Science of the U.S. Department of Energy
under Contract No. DE-AC02-05CH11231. Computations were performed at
\emph{Carver} and \emph{Edison}.

\appendix

\section{Neutrino Treatment}
\label{s:neutrino_details}

\subsection{Neutrino Leakage Scheme and Boundary Layer}

Following \citet{ruffert1996}, we treat neutrino cooling by smoothly interpolating
between the direct neutrino loss rate in the optically thin limit, and the
diffusive loss rate of neutrinos in chemical equilibrium for large optical depth.
We do this because we reach optical depths $\gtrsim 10$ in the boundary layer
for an extended period of time.

The effective rate per baryon of lepton number loss and the rate per unit mass of 
energy loss are given respectively by
\begin{eqnarray}
\Gamma^{\rm eff}_i & = & \frac{\Gamma^0_i}{1 + t_{{\rm n-loss},i}/t_{{\rm diff},i}}\\
Q^{\rm eff}_i      & = & \frac{Q^0_i}{1 + t_{{\rm e-loss},i}/t_{{\rm diff},i}},
\end{eqnarray}
where the subscript $i$ refers to electron-type neutrinos or antineutrinos, 
$t_{{\rm diff},i}$ is the diffusion time, and the direct loss timescales satisfy
\begin{eqnarray}
t_{{\rm n-loss},i} & = & Y_i/\Gamma^0_i\\
t_{{\rm e-loss},i} & = & e_i/Q^0_i,
\end{eqnarray}
with $\Gamma^0_i$ and $Q^0_i$ the optically-thin lepton number and energy loss rates,
respectively, $Y_i$ the neutrino number per baryon in chemical equilibrium, and $e_i$ the
specific neutrino energy in chemical equilibrium \citep{ruffert1996}.

The diffusion time for each neutrino species is given by $t_{{\rm diff},i} = (L/c)\tau_i$,
where $L$ is a characteristic distance, $c$ is the speed of light, and $\tau_i$ is the
optical depth of species $i$ over the distance $L$. To keep computations economical,
we define the optical depth as
\begin{equation}
\label{eq:tau_hp}
\tau_i = \kappa_i \min{\left\{r,H_\perp,H_\parallel\right\}},
\end{equation}
where $\kappa_i$ is the corresponding charged-current absorption coefficient, and 
$H_\perp$ and $H_\parallel$ are the vertical and horizontal gas pressure
scale heights, respectively, obtained using the appropriate components of the effective gravitational
acceleration vector $\mathbf{g}_{\rm eff}$ (e.g., 
$H_\perp = p / (\rho |\mathbf{g}_{\rm eff}| \sin\theta$). This prescription approximates
the true optical depth to within a factor of a few. The same length scale used in 
equation~(\ref{eq:tau_hp}) is assigned to $L$ for computing the diffusion time $t_{{\rm diff},i}$.

\begin{figure}
\includegraphics*[width=\columnwidth]{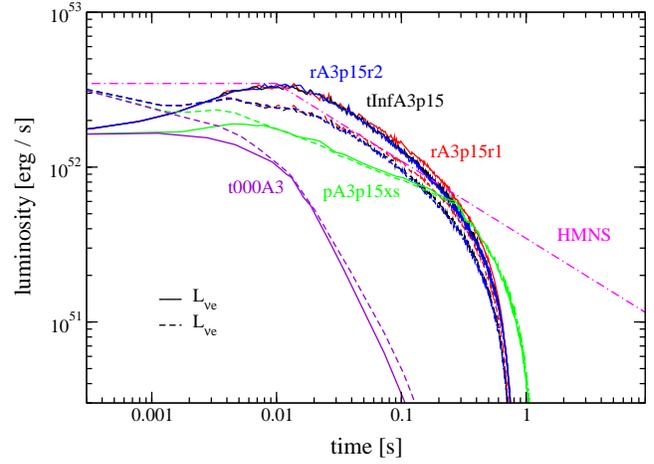}
\caption{Neutrino and antineutrino luminosities from the disk as a function of time (solid and
dashed lines, respectively). Shown are three models with different radial
resolutions near the inner boundary (rA3p15r1, tInfA3p15, rA3p15r2), a model
with irradiation from the HMNS suppressed, keeping the reflecting boundary and disk
self-irradiation (pA3p15xs), and a model with a BH at the center (t000A3).
The imposed HMNS luminosity (equal for neutrinos and antineutrinos) is shown
by the dot-dashed magenta line.}
\label{f:luminosity_convergence}
\end{figure}

Figure~\ref{f:luminosity_convergence} shows the resulting neutrino luminosities
from the disk for different models. Given that at late times most of the emission
arises from a thin boundary layer, we have tested the sensitivity of the
resulting luminosity to the spatial resolution. Models rA3p15r1 and rA3p15r2
have inner radial cells twice larger and smaller than the fiducial model
tInfA3p15, respectively\footnote{Since the radial grid is ratioed, this means that
the increased resolution in inner regions is compensated by lower resolution outwards for
a fixed number of cells.}. 
The luminosities are very close to each other, with an 
overall trend of less emission with higher resolution. Given this outcome we consider
the neutrino emission to be converged. The thermodynamic quantities of the
ejecta differ by a few percent or less between models with different
resolution. The ejected mass along the equator is the most sensitive,
with differences up to a few percent. Model rA3p15r2 is terminated
early due to the higher computational cost; comparing the mass outflow
rate with that of tInfA3p15 at the same time yields differences 
up to $\sim 10\%$ in equatorial ejecta mass.

The effect of the HMNS irradiation on the disk emissivity can be 
assessed by comparing the luminosity from model pA3p15xs with that of 
the fiducial model tInfA3p15. When the HMNS emits, the disk luminosity
increases relative to no HMNS irradiation over the thermal time of the
disk ($\sim 30$~ms at $r\sim 100$~km). The output luminosity exceeds the imposed HMNS
flux due to the contribution from accretion.

The effect of the reflecting boundary can be quantified by comparing
the luminosity of model pA3p15xs with that of model t000A3. The increased
radiative efficiency results in higher emission by about an order
of magnitude at time $\sim 0.1$s, explaining why both neutrino
energy deposition and composition changes are more prominent
when the HMNS is present.

\subsection{Self-Irradiation with a Boundary Layer}

The spatial emission properties of the disk are slightly more complicated when
a HMNS sits at the center than when a BH is the central object. Most of
the emission comes from two hot spots located above and below the
optically-thick midplane (Fig.~\ref{f:2d_map}c). Thus our prescription for \emph{disk
self-irradiation} requires modifications relative to FM13.  Since the position
of these hot spots evolves in both radius and polar angle, we compute the
neutrino temperature as an emissivity-weighted average over the entire
computational domain instead of a mass-weighted average.  This way we avoid
biasing this temperature towards the optically thick midplane, which does not
emit much but has the highest temperature. We ignore differences between
neutrino and antineutrino temperature from the disk.   

Given that the emission is equatorially symmetric, we use the same angular
distribution as in FM13, with the emission radius $R_{\rm em}$ computed in the
same way.  However, the attenuation of the radiation flux from the disk is
modified in two ways. First, we multiply by an overall factor of 1/2 to account
for the fact that the midplane acts as a radiation insulator. Second, we use an
attenuation optical depth consistent with that employed in the leakage
calculation. That is, the radiation flux from the disk seen by a given point in
the computational domain is multiplied by a factor $\exp{(-\tau_{{\rm
irr},i})}$, with $\tau_{{\rm irr}}$ being the maximum of the values
obtained by evaluating  equation~(\ref{eq:tau_hp}) at the location of the
emission hot spot and at the absorption point.  This
prescription is good at capturing large density contrasts, but fails to account
for shadowing at large radii along the equator.

\bibliographystyle{mn2e}
\bibliography{ms}

\label{lastpage}
\end{document}